\documentclass[prb,twocolumn,superscriptaddress]{revtex4-1}
\usepackage{amsmath,amssymb,bm}
\usepackage{hyperref}
\usepackage{graphicx}
\usepackage{epstopdf}
\usepackage{latexsym}
\usepackage[usenames, dvipsnames]{color}
\usepackage[usenames, dvipsnames]{xcolor}
\usepackage{natbib}
\usepackage{braket}
\usepackage{float}
\usepackage[normalem]{ulem}
\usepackage{comment}
\usepackage{mathtools}
\usepackage{array}
\usepackage{tabu}
\usepackage{multirow}
\usepackage{bbold}
\usepackage{commath}
\usepackage{lipsum}

\begin{document}
\title{Layer Coherent Phase in Double Layer graphene at $\nu^{}_1=\nu^{}_2=0$}
\author{Amartya Saha}
\email{amartya.saha@uky.edu}
\affiliation{Department of Physics and Astronomy, University of Kentucky,
Lexington, KY 40506, USA}
\author{Ankur Das}
\email{ankur.das@weizmann.ac.il}
\affiliation{Department of Condensed Matter Physics, Weizmann Institute
of Science, Rehovot, 76100 Israel}

\begin{abstract}
In the recent advancement in graphene heterostructures, it is possible to
create a double layer tunnel decoupled graphene system that has a strong
interlayer electronic interaction. In this work, we restrict the parameters
in the low energy effective Hamiltonian using simple symmetry arguments.
Then, we study the ground state of this system in the Hartree-Fock
approximation at $\nu^{}_1=\nu^{}_2=0$. In addition to the phases found in
monolayer graphene, we found an existence of layer coherent phase which
breaks the layer $U(1)$ symmetry. At non-zero Zeeman coupling strength
($E^{}_z$), this layer coherent state has a small magnetization, that
vanishes when $E^{}_z$ tends to zero. We discuss the bulk gapless modes
using the Goldstone theorem. We also comment on the edge structure for the
layer coherent phase. 
\end{abstract}

\maketitle

\section{Introduction} After the discovery of the quantum Hall effect in
two dimensional electron gas (2DEG) \cite{Klitzing_1980}, the interest
started to build towards bilayer 2DEG's \cite{Chakraborty_Pekka_1987}
(and for general systems with internal quantum number \cite{Halperin_1983}).
In the quantum Hall regime the ground state of double layer 2DEG  has been
explored theoretically \cite{jain_2007,Scarola_Jain_2001,Chakraborty_Pekka_1987}
and later experimentally \cite{Eisenstein_2014} (and the references therein).
It has been observed that in 2DEG (in this case GaAs), when the interlayer distance
is small, the system at total filling fraction $\nu^{}_T=\nu^{}_1+\nu^{}_2=\frac{1}{2}$
($\nu^{}_1$, $\nu^{}_2$ being the filling fraction of each layer) and $1$ forms an
incompressible QHS\cite{Kellogg_etal_2004,Gramila_etal_1991,Eisenstein_MacDonald_2004}.
The individual layers at $\nu^{}_T=1$ and $\frac{1}{2}$ has even denominator filling
fractions of $\frac{1}{2}$ and $\frac{1}{4}$ respectively which are known to be
compressible states. This phase is a layer coherent phase in which the electron of one 
layer forms a bound state with the
holes of the other layer forming excitons and sponteneously breaks the layer $U(1)$
symmetry. We can think of layer coherent states as either
easy plane layer pseudospin ferromagnet or electron-hole bound exciton
\cite{Eisenstein_2014}.
This arises from the conservation of particle number in the individual layers.
Developments in graphene technology have boosted the interest in quantum Hall effect
in graphene \cite{Novoselov_etal_2005,Zhang_etal_2005}. Some recent experiments in the
quantum Hall regime in double layer  graphene systems has been found to exhibit layer
coherent states\cite{Li_etal_2016,Li_etal_2019,Liu_etal_2019}. Experimentally it is
possible to fabricate double layer graphene with very small interlayer distance $d$
($\sim$ 2nm) where $d/l<1$ ($l$ being the magnetic length) \cite{Gorbachev_etal_2012,Titov_etal_2013,Li_etal_2016,Liu_etal_2017,Li_etal_2017,Britnell_etal_2012,Lee_etal_2016,Liu_etal_2017, Liu_etal_2019,Li_etal_2019} which was earlier
difficult to achieve in GaAs systems. The separator between the graphene layers is
made out of stacked hBN layers. Thus by changing the number of stacked hBN layers
the interlayer interaction can be tuned from weak to strong. This
induced a huge interest in the understanding and testing of the double layers
of graphene, Bernal-stacked bilayer graphene \cite{Li_etal_2016,Li_etal_2017},
twisted magic angle bilayer graphene \cite{Lui_etal_2020_TBG} etc. There has
been some theoretical \cite{Tse_Hu_Das_2007,Katsnelson_2011,Peres_Santos_Neto_2011,Narozhny_etal_2012, Kharitonov_2010}
and experimental \cite{Kim_etal_2011} studies to understand the Coulomb drag
in double layer graphene in zero magnetic field. Ref.
\onlinecite{Hongki_etal_2008} predicted that at higher temperature at the zero
magnetic field there can be a normal to superfluid transition in double layer
graphene.

In the presence of a ultra short-range (compared to the magnetic length $l$)
interaction, the Hamiltonian projected to the $n=0$ Landau level manifold for
monolayer graphene (MLG) has $SU(2)_\text{spin}\times\left(U(1)\times
\mathbb{Z}_2 \right)_\text{valley}$ symmetry in the absence of a Zeeman
field \cite{Alicea_Fisher_2006}. There exists four different
possible phases namely ferromagnet (F), charge density wave (CDW),
Kekul\'e distorted phase (KD) and anti-ferromagnet (AF), which becomes
canted anti-ferromagnet (CAF) in presence of Zeeman coupling
\cite{Kharitonov_2012}. The predicted phase transition from CAF to F
\cite{Herbut1,Herbut2,Kharitonov_2012_MonoEdge,Murthy_Shimshoni_Fertig_2014, Murthy_Shimshoni_Fertig_2016,Knothe_Jolicoeur_2015} has been verified
in the experiment \cite{Young_etal_2014}. This understanding of symmetry
has been used to study the ground state at fractional fillings as well
\cite{Sodemann_MacDonald_2014}.

In the case of double monolayer graphene electron fillings of each
layer ($\nu^{}_1$ and $\nu^{}_2$) can be controlled independently. For
this system many interesting quantum states have been found. Some
of those states can be explained using interlayer Jain composite fermion
states \cite{jain_2007}, proposed for double layer two-dimensional electron
gas \cite{Li_etal_2019}. In this manuscript we propose the relevant symmetry
in double monolayer graphene which restricts the interacting Hamiltonian to
three parameters. Within the scope of this letter, we restrict
ourselves to understand the mean field ground state when two layers of
graphene are at $\nu^{}_1=\nu^{}_2=0$. We show that for certain values of the 
parameters, the system goes into a layer coherent phase, which has a small 
magnetization in presence of a Zeeman field. Increasing the Zeeman field strength
one can drive a second order phase transition from magnetized layer coherent phase
to the ferromagnetic phase.

Here we would like to emphasize that we want to find a low energy
Hamiltonian that is restricted by symmetry. We also focus on the translation
invariant ground state solutions of this low energy
Hamiltonian in the allowed parameter space. Our method does not talk
about the details of the microscopic model but only the low energy effective
model. 

We describe our assumptions, method and findings in a few sections.
In section \ref{sec:assum} we describe the assumptions and our Hamiltonian. 
After that we describe the results and the Goldstone modes in section
\ref{sec:results} and \ref{sec:gold} respectively. We also have some
discussion over possible lattice models, experimental signatures in section
\ref{sec:discussion}. Then in section \ref{sec:summary} we summarize 
our findings and describe possible application of this work.

\section{Assumptions and Model}
\label{sec:assum}
We restrict our calculations to the $n=0$ Landau level. When the interaction strength
is much smaller than the cyclotron energy gap the Landau level mixing can be ignored.
In the strong interaction strength regime the form of the effective theory gets
dictated by the symmetry (discussed below) when we integrate out
the higher Landau levels. When the layers are far enough from each other, we
should get two MLG with no interlayer interaction. The valley $U(1)$ for 
each layer is conserved in order to conserve 
the translational symmetry in each layer separately.
Here we make an additional
assumption that the global spin $SU(2)$ symmetry can be enhanced to spin $SU(2)$ 
symmetry for each layer separately. For this to be the symmetry of this theory
we assume that interlayer spin-spin interaction is zero (or negligible). In the
absence of the inter layer tunneling it is justified that the Heisenberg term 
($\vec{S}\cdot\vec{S}$) will be absent. Other than the Heisenberg a long range
spin dipole-dipole interaction between layers can break the spin $SU(2)$ symmetry
in each layer to a global $SU(2)$ symmetry. However, the spin dipole-dipole interaction
falls as $r^{-4}$. As the distance between the layers is a few nanometers, we
choose to ignore this interaction. Thus under these assumptions the only term that is
allowed is the $\vec{S}\cdot \vec{S}$ on each layer which has the spin $SU(2)$ symmetry
in each layer.

From this understanding and keeping in mind that the number of particles in each
layer is fixed we propose our symmetry of the continuum model to be (in the
absence of Zeeman coupling)
$\left[SU(2)_\text{spin}\otimes U(1)_\text{valley}\right]$
for each layer, a global $\left(\mathbb{Z}_2\right)_\text{valley}$ and
$\left( U(1) \otimes \mathbb{Z}_2\right)_\text{layer}$ for the
layers.
symetry for each layer.This restricts the interacting part of the Hamiltonian 
to only three parameters. We can write the Hamiltonian as,

\begin{align}
H = H_0 + H_\text{int} .
\end{align}
where $H_0$ is the one body term coming from Zeeman coupling such that,
\begin{align}
H_0 = -E^{}_z \left( \sigma^z \otimes \tau^0 \otimes \gamma^0 \right) .
\end{align}


$H_\text{int}$, the 2 body interaction term which obeys the above mentioned
symmetry, is given by
\onecolumngrid

\begin{align}
\nonumber
H_{int}=\frac{\pi l^2}{A}\sum_{\substack{\vec{q}\\k_1,k_2}} e^{-iq_x(k_1-k_2-q_y)l^2-\frac{q^2l^2}{2}} &\Bigg[ K_{xy} \sum_{i=1,2}:\vec{c}^\dagger_{k_1-q_y}\left(\sigma^0 \otimes \tau^i \otimes P_L\right)\vec{c}^{\phantom{\dagger}}_{k_1}\vec{c}^\dagger_{k_2+q_y}\left(\sigma^0 \otimes \tau^i \otimes P_L\right)\vec{c}^{\phantom{\dagger}}_{k_2}: \\
\label{eq:contham}
& + K_z:\vec{c}^\dagger_{k_1-q_y}\left(\sigma^0 \otimes \tau^3 \otimes P_L\right)\vec{c}^{\phantom{\dagger}}_{k_1}\vec{c}^\dagger_{k_2+q_y}\left(\sigma^0 \otimes \tau^3 \otimes P_L\right)\vec{c}^{\phantom{\dagger}}_{k_2}: \\ \nonumber
& + L_z:\vec{c}^\dagger_{k_1-q_y}\left(\sigma^0
\otimes \tau^0 \otimes \gamma^3\right)\vec{c}^{\phantom{\dagger}}_{k_1}\vec{c}^\dagger_{k_2+q_y}\left(\sigma^0 \otimes \tau^0 \otimes \gamma^3\right)\vec{c}^{\phantom{\dagger}}_{k_2}:\Bigg]
\end{align}

\twocolumngrid

Here $\vec{c}_k=\big(c^{\phantom{\dagger}}_{k,\uparrow,K,1}, c^{\phantom{\dagger}}_{k,\downarrow,K,1},
c^{\phantom{\dagger}}_{k,\uparrow,K',1}, c^{\phantom{\dagger}}_{k,\downarrow,K',1},
c^{\phantom{\dagger}}_{k,\uparrow,K,2}, \allowbreak c^{\phantom{\dagger}}_{k,\downarrow,K,2},
\allowbreak c^{\phantom{\dagger}}_{k,\uparrow,K',2}, c^{\phantom{\dagger}}_{k,\downarrow,K',2} \big)^T$
presents the column vector of fermionic annihilation operators, $A$ is the area of the
sample, and $l$ is the magnetic length. The index $k_i$ represents the guiding centers
in the Landau gauge.
We use the convention where $\sigma^i, \tau^i, \gamma^i$ represents the
Pauli matrices in spin, valley, and layer respectively. Here,
$P_{L}=\frac{\left( \gamma^0 - \left(-1\right)^L \gamma^3 \right)}{2}$ is the layer
projection operator to layer $L$. The parameters $K_z$ and $K_{xy}$ arises from the
intralayer interactions and are same as the parameters $u_z$ and $u_{\perp}$
respectively as defined by Kharitonov in the monolayer graphene
case\cite{Kharitonov_2010}. The parameter $L_z$ is a function of the distance between
the layers ($d$) which should go to zero as $d$ becomes very large (disjoint MLG limit).
Here we would like to comment that we also added a capacitance term which is zero
when both layers have equal fillings \cite{Wen_Zee_1992},
\begin{equation}
H_{cap}=\frac{g_{es}\pi l^2}{A}\left(\rho_{1}(\mathbf{q}=0)-\rho_{2}(\mathbf{q}=0)\right)^2.
\end{equation}
where $\rho_{L}(\vec{q})$ is the Fourier transformed electron density operator of $L$th
layer and $g_{es}$ is the coupling strength of the capacitance term. 

We define an order parameter $\Delta$ matrix which specifies the HF states $|HF\rangle$,
\begin{equation}
\langle HF|c^\dagger_{k,s,\alpha,L} c^{\phantom{\dagger}}_{k,s,'\alpha', L'}|HF\rangle=\delta_{\vec{k},\vec{k}'}\Delta^{}_{s'\alpha' L',s\alpha L}.
\end{equation}
where $s$ being the spin, $\alpha$ being the valley and $L$ being the layer index.
This $\Delta$ matrix can also be thought of as a sum of projection
operators of the four filled states at each momemtum.
The $\Delta$ matrix completely determines the single Slater determinant states
and any order parameters e.g. electron density, magnetization etc. 
can be calculated using it. 
We assume that the HF states preserve translation symmetry i.e. the guiding
centers are a good quantum number. Hence, we drop the guiding center label from
$\Delta$ matrix. Since the capacitance term is a classical term, we only keep
the Hartree term and drop the Fock term. In the next section, we discuss the
$\Delta$ matrix of the different HF states.

\begin{figure}
\centering
\includegraphics[width=\columnwidth]{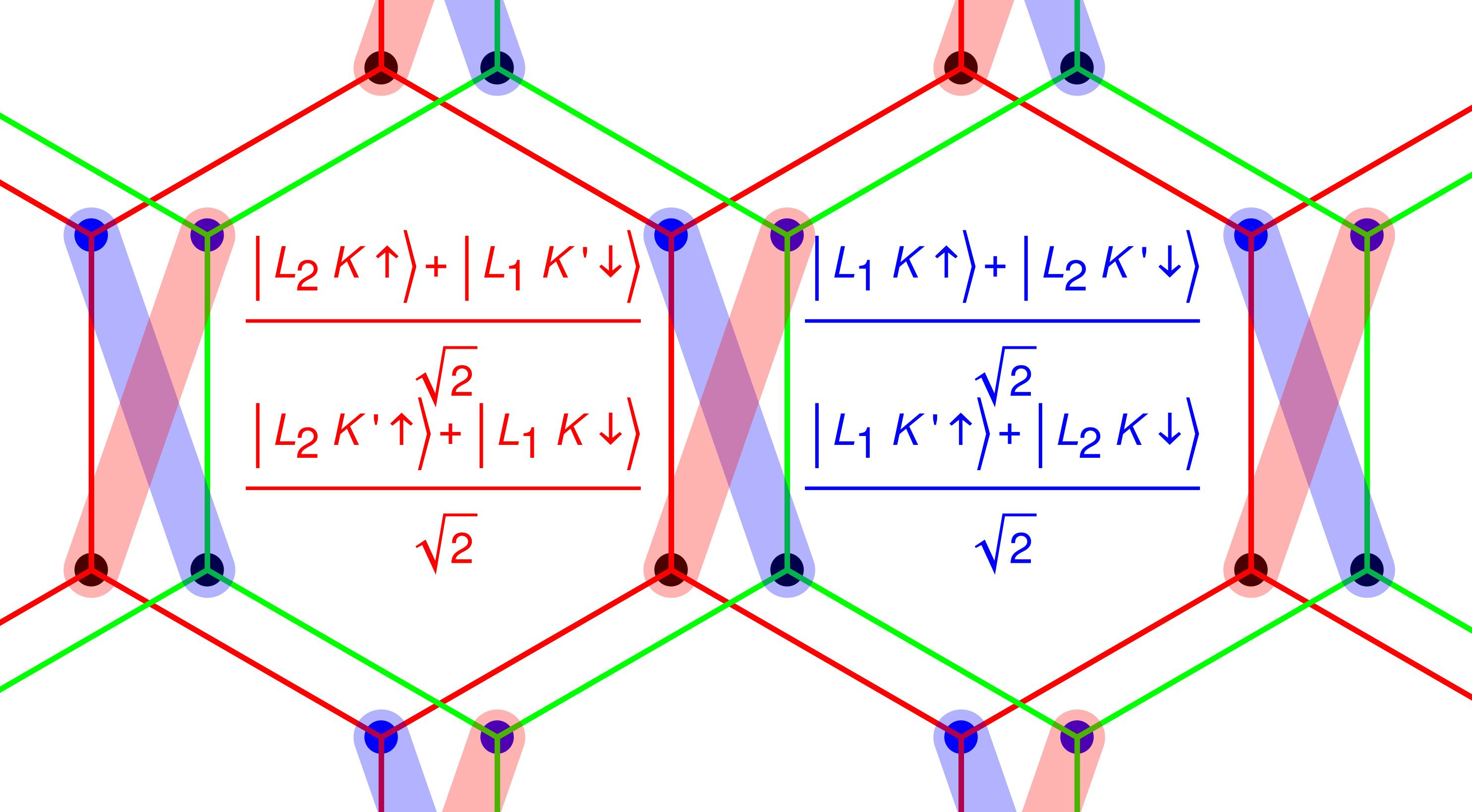}
\caption{Here we represented the Layer coherent (LC) phase. The layers here are color
coded (green lines and red lines). As shown here the states are linear combination
of different layer indices. The corresponding states are also color coded as blue and
light red.}
\label{fig:LCphase}
\end{figure}

\section{Results}
\label{sec:results}

At $\nu^{}_1=\nu^{}_2=0$ there are four occupied single particle states in the
spin-valley-layer space. For $L_z\geq 0$ we find that the phase diagram is
exactly the same as the phase diagram found for MLG in Ref. \onlinecite{Kharitonov_2012}.
The energies of the phases
(defined as $E_{gs}=\langle HF|H_\text{int}|HF \rangle$ for the proposed
HF ground state)) depend on $L_z$. For these phases, the layer $U(1)$ is not broken
and the $\Delta$ matrix is block diagonal in the
layer index. Each of this block is a four dimensional matrix in the space of
valley and spin. The four phases are,
\begin{enumerate}
 \item Charge Density Wave (CDW): CDW breaks the valley $\mathbb{Z}_2$
 symmetry. At the zero Landau level, different valley indices are pinned to
 the sublattices. In this phase in each layer the alternate sites (A)
 in the lattice are occupied and the other sites (B) are left unoccupied.
 The $\Delta$ matrix for this phase is
\begin{equation}
 \Delta_\text{CDW}=\sigma^0 \otimes (\tau^0+\tau^3) \otimes \gamma^0,
\end{equation}
and the energy is
\begin{equation}
E_\text{CDW}=2\left(K_z-L_Z\right). 
\end{equation}
 \item Kekul\'e Distorted (KD): This is a bond order phase where the valley $U(1)$
 symmetry is broken. In lattice limit the spontaneous breaking of the valley $U(1)$
 symmetry leads to the translation symmetry breaking in each layer. This phase
 doesn't has any Goldstone modes. The $\Delta$ matrix for this phase will be
\begin{equation}
 \Delta_\text{KD}=\frac{1}{2}\sigma^0 \otimes (\tau^0+\tau^1) \otimes \gamma^0,
\end{equation}
with energy,
\begin{equation}
E_\text{KD}=2\left( K_{xy} -L_z\right). 
\end{equation}
 \item Ferromagnet (F): This phase breaks the spin $SU(2)$ symmetry in each layer.
 Similarly, the $\Delta$ matrix and energy will be
\begin{align}
\Delta_\text{F}&=\frac{1}{2}(\sigma^0+\sigma^3) \otimes \tau^0 \otimes \gamma^0\\
E_\text{F}&=-4E^{}_z-2\left( 2K_{xy}+K_z+L_Z \right).  
\end{align}
 \item Canted Anti-Ferromagnet (CAF): This phase breaks the spin $U(1)$ symmetry
 in each layer. The $\Delta$ matrix is,
\begin{align}
\Delta_\text{CAF}&=\frac{1}{2}\Big[\sin\phi \left(\sigma^1 \otimes \tau^3 \otimes \gamma^0\right)\nonumber \\
&+\cos\phi \left(\sigma^3 \otimes \tau^0 \otimes \gamma^0 \right) + \sigma^0 \otimes \tau^0\otimes \gamma^0\Big]
\label{eq:Double_CAF}
\end{align}
where $\phi$ is given by
\begin{equation}
\cos\phi=\frac{E^{}_z}{2 |{K_{xy}}|},
\end{equation}
\end{enumerate}
with energy,
\begin{equation}
E_\text{CAF}=-4E^{}_z-2\left( 2K_{xy}+K_z+L_Z \right). 
\end{equation}
At $E^{}_z=0$ the states becomes a pure anti-ferromagnetic state. 
Increasing the Zeeman field
$E_z$ beyond $2|K_{xy}|$ can drive a continuous phase transition from canted 
anti-ferromagnetic phase to ferromagnetic phase.

\begin{figure}[h]
\centering
\includegraphics[width=\columnwidth]{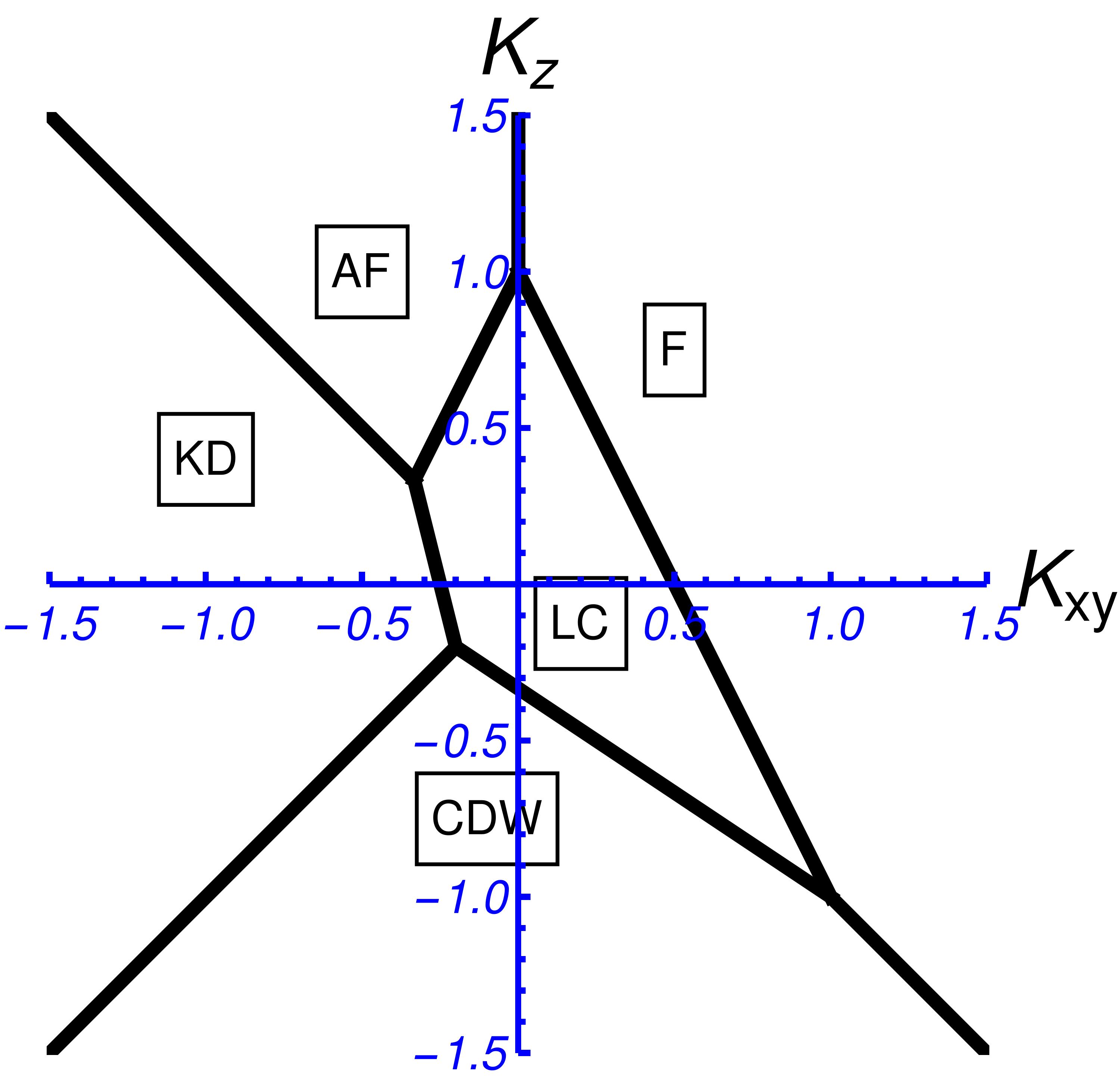}
\caption{Here we present the Phase diagram of for $L_z=-0.5$ in the absence of
Zeeman coupling. The phase LC Phase appears near $K_{xy}=K_z=0$. All the phase
transitions here are First order.}
\label{fig:negLzPhaseDiagnoEz}
\end{figure}

\begin{figure}[h]
\centering
\includegraphics[width=\columnwidth]{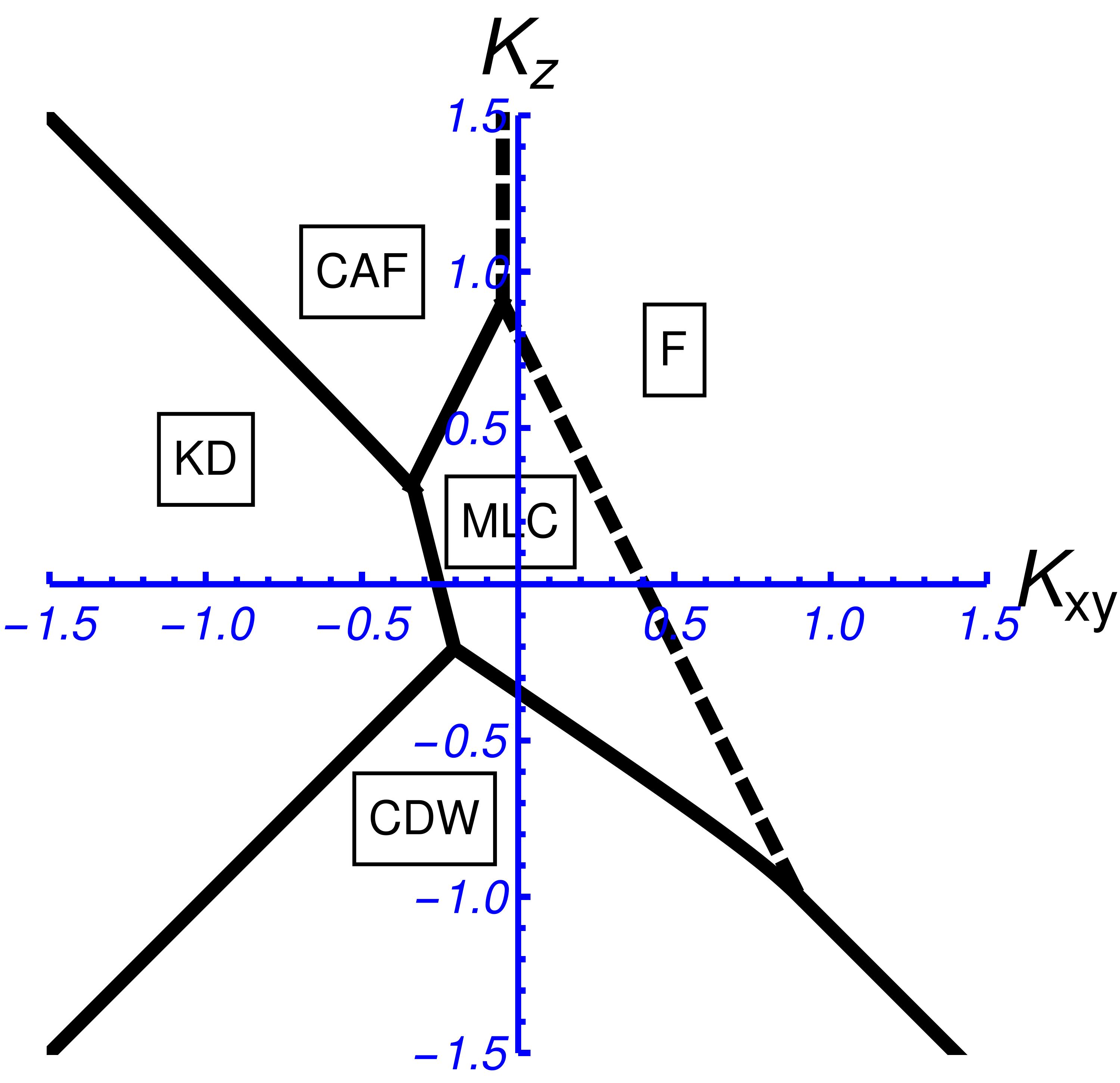}
\caption{Here we present the Phase diagram of for $L_z=-0.5$
and $E^{}_z=0.1$. As we can see that the MLC phase
appears and there is a second order transition from MLC to F
as marked by the broken line.}
\label{fig:negLzPhaseDiag}
\end{figure}

Next we come to the spacial phase of the double layer graphene. For $L_z<0$,
we find there exists a layer coherent phase which breaks the Layer $U(1)$
symmetry  (see \ref{fig:negLzPhaseDiag}). We find the layer coherent phase
both in the presence and absence of the Zeeman energy. For a non-zero Zeeman
coupling, there are two parameters (and operators which are connected by the
left over ground state symmetry) that,
\begin{subequations}
\begin{align}
\Phi_L=\sigma^1 \times \tau^1 \times \gamma^1\\
S^z=\frac{\sigma^3 \times \tau^0 \times \gamma^0}{2}.
\end{align}
\end{subequations}

The magnetic layer coherent phase occurs at $E^{}_z \neq 0$ with
$\langle S^z\rangle\neq 0$. At $E^{}_z=0$ we find $\langle S^z \rangle= 0$,
we call it layer coherent phase (LC) (see fig. \ref{fig:LCphase}). We can write
the $\Delta$ matrix for the MLC phase as,
\begin{align}
\Delta_\text{MLC}=&\frac{1}{2}\Big[\sin\theta (\sigma^1 \otimes \tau^1 \otimes \gamma^1) \nonumber\\
&+ \cos\theta (\sigma^3 \otimes \tau^0 \otimes \gamma^0)+ \sigma^0 \otimes \tau^0\otimes \gamma^0\Big]
\label{eq:layerOrder}
\end{align}
with  $\cos\theta$ defined as,
\begin{equation}
\cos\theta=\frac{2 E^{}_z}{\abs{2 K_{xy}+K_z+2 L_z}}.
\end{equation}
The energy of the phase is 
\begin{align}
E_\text{MLC}=-2K_{xy}-K_z - \frac{4E^2_z}{\abs{2K_{xy} +K_z +2L_z}}.
\end{align}
Here $\langle\Phi_L\rangle=4\sin\theta$ and $\langle S^z\rangle=
2 \cos\theta$. For $2 E^{}_z\geq \abs{2 K_{xy}+K_z+2 L_z}$ with $\theta = 0$, this
$\Delta$ will represent a ferromagnetic ground state and for any other value of
$\theta$ the $\Delta$ matrix represents a magnetic layer coherent state (MLC).
For the zero
Zeeman coupling, we have $\cos \theta=1 \Rightarrow \langle S^z\rangle=0$, a
purely layer coherent state. The phase transition from MLC to F is a second
order transition (see Fig. \ref{fig:negLzPhaseDiag}). However, similar to the
AF to F phase transition, the LC to F phase transition is a first order
transition. Thus all phase transitions are first order at $E^{}_z=0$ (see Fig.
\ref{fig:negLzPhaseDiagnoEz}). The Phase boundary between MLC and F changes, as
we change the total Zeeman couplings at a fixed $L_z$. Here in Table
\ref{table:Double_boundary} we represent all different phase boundaries.

\begin{table}[H]
\centering
\begin{tabular}{ ||c||c|| }
\hline
Phases & Boundary equation\\
\hline
\hline
KD,CAF & $K_z=-K_{xy}+E^2_z/K_{xy}$ \\
\hline
KD,CDW & $K_z=K_{xy}$\\
\hline
F,CDW & $K_z=-K_{xy}+E^{}_z$\\
\hline
F,MLC & $K_z=-2K_{xy}-2L_z-2E^{}_z$\\
\hline
MLC,CDW & $-\frac{3 K_z}{2}=\left(L_z+2K_{xy}-\sqrt{\left(2L_z+K_{xy}\right)^2+3E^2_z}\right)$\\
\hline
CAF,MLC & $K_z=2\left(K_{xy}-L_z\right)$ \\
\hline
KD,MLC & $K_z=-3K_{xy}-\sqrt{\left(K_{xy}-2L_z\right)^2+4E^2_z}$\\
\hline
F,CAF & $K_{xy}=-E^{}_z/2$\\
\hline
\end{tabular}
\caption{Phase boundary equations as a function of the parameters}
\label{table:Double_boundary}
\end{table}

\section{Goldstone modes}
\label{sec:gold}
The Hamiltonian in the presence of Zeeman term has five different $U(1)$
symmetries coming from $U(1)_\text{spin} \otimes U(1)_\text{valley}$
for the two layers and a layer $U(1) \times \mathbb{Z}_2$ symmetry. For the layer
diagonal phases, the presence of gapless bulk Goldstone mode is known.
The CDW phase has no gapless bulk mode. In the continuum limits it seems that
KD phase breaks a continuous symmetry but as valley indices are momenta, it
breaks lattice symmetry. Hence in this phase we will have no Goldstone
modes. The F phase has spin wave mode and at long wavelength, it's gap is
proportional to the Zeeman coupling strength ($E^{}_z$). As the CAF phase
breaks the spin $U(1)$ symmetry there will be a pair of gapless neutral modes
in the bulk \cite{Murthy_Shimshoni_Fertig_2014}.

Next we discuss the new layer coherent phase and its bulk modes. From Eq.
\ref{eq:layerOrder} one can easily see that the ground state has the two
leftover $U(1)$ symmetries defined by operators $\sigma^3 \otimes \tau^3
\otimes \gamma^0$ and $\sigma^3 \otimes \tau^0 \otimes \gamma^3$. These
operations can be understood as opposite spin rotations at different valleys
or different layers. In other words, these are relative valley and layer
spin twists respectively. Thus out of five different continuous symmetries,
three are broken by the ground states giving rise to three different
Goldstone modes in the bulk. However, these modes will be neutral as there
is a charge gap in the bulk and these excitations are similar to spin waves.
We remind the readers here that the breaking of the valley part of the
symmetry breaks the lattice $\mathcal{C}_3$ rotation about a site. This
happens as the $n=0$ manifold the $\mathbf{K}, \mathbf{K}'$ of each layer
maps to the $A,B$ sublattice of each layer\cite{Kharitonov_2012}. This means
we will count one extra Goldstone mode in the continuum analysis.


\section{Discussion}
\label{sec:discussion}
In this manuscript, we constructed the Hamiltonian using symmetry principles
without discussing the nature and details of the interaction at the lattice scale.
The model only assumes the lattice interactions are local and thus their Fourier
transform is a function independent of momentum.

In principle we can reproduce the interactions in the continuum model by projecting
the microscopic Hamiltonian to the lowest Landau Level. We present a simplified
example which includes the
onsite Hubbard interaction ($U_1$), nearest neighbor interlayer electron
density-density interaction ($U_2$) and a intralayer nearest neighbor spin-spin
interaction ($J$),
\begin{align}
H_\text{lat}=&U_1\sum_{\substack{s_1,s_2\\r,L}} :n_{s_1,r,L} n_{s_2,r,L}: + U_2
\sum_{\substack{s_1,s_2\\r,r'\\L_1\neq L_2}} :n_{s_1,r,L_1} n_{s_2,r',L_2}:\nonumber \\
&+J\sum_{\substack{\langle r,r' \rangle,L}} :\vec{S}_{r,L} \cdot \vec{S}_{r',L}:.
\label{eq:latHam}
\end{align}
Here $n$ is the fermion number operator and $\vec{S}$ is the local spin operator.
Though we assumed ultra-short interactions, adding finite
range to these interactions do not change the symmetry of the continuum model when
we project the hamilitonian in the zero Landau level manifold. We can write
the relation between the continuum parameters in Eq. \ref{eq:contham} in terms of
the parameters of Eq. \ref{eq:latHam} as $K_{xy}\propto -J$, $K_z\propto U_1$ and
$L_z\propto U_1/2-U_2$.

We would like to emphasise here that the Hamiltonian presented here is
just an example to show that even at the simplest model at the lattice
level, we can achieve the Hamiltonian in Eq. \ref{eq:contham}. Here we are
not concerned with the values/signs of different parameters $K_{xy}, K_z, L_z$ but
showing the phases that is determined by these parameters.
Modeling a generic lattice theory with physically motivated parameters and their values
is an interesting study but out of the scope for the current manuscript.
We hope to study a lattice model of a double layer graphene in the future.

From Eq. \ref{eq:layerOrder} we can see that as the states are mixture of
$\vec{K},\vec{K}'$ of different layers, near the edge the dispersion will
contain two pairs of particle-like bands and another two pairs of hole-like
bands due to the breaking of the
translation symmetry \cite{Murthy_Shimshoni_Fertig_2017}. There will be a pair of
counter propagating modes only if we have identical layers near the edge of the system.
Near the edge,
both the valley $U(1)$ symmetry and the layer $U(1)$ will be broken generically.
Thus the edge of a double layer graphene will be gapped if the bulk is in MLC phase.
As the states
are superposition of two different layers, there will be a drag in the two terminal
measurements at least for finite temperature \cite{Titov_etal_2013}.
However, we know that there will be
two pair of counter propagating modes at the edge for each layer in the F phase
\cite{Kharitonov_2012_MonoEdge,Murthy_Shimshoni_Fertig_2017,Young_etal_2014}.
Thus by changing the Zeeman energy with respect to the interaction
energies, one can make a transition from MLC to F. This should show up in the
two terminal conductance measurement\cite{Young_etal_2014}. It will also be interesting
to measure the lattice scale structure using both spin resolved and spin unresolved
\cite{Yu_etal_2019} tunneling electron microscope to confirm the phase directly.

\section{Summary and outlook}
\label{sec:summary}
Here we argued that the continuum limit of the double layered graphene at
$\nu^{}_1=\nu^{}_2=0$ has a big symmetry group that restricts the interacting part
of the Hamiltonian severely to only three parameters at $n=0$ Landau Level.
Further, we find a candidate ground state using the HF approximation that
breaks the layer $U(1)$ symmetry. We also find a second order phase transition
from MLC to F as a function of Zeeman energy. We argued for a general system,
the edge of the MLC will be gapped. This leads to the possible experiment to
find two terminal conductance that will change when we go from the MLC to F. 

This study is just the beginning of understanding the double layer graphene
ground state in the quantum Hall regime. It was previously shown that the phase
transition from CAF to F connects the bulk gapless modes of the CAF to the
gapless edge modes of F \cite{Murthy_Shimshoni_Fertig_2014}. We hope to study
the edge theory of double layer graphene in future to answer the question of
the phase transition from MLC to F.


It has been shown that if we have finite range interactions in monolayer
graphene then we can have co-existence of phases
\cite{Das_Kaul_Murthy_2021}. Similarly, in a lattice model co-existence can also be
shown by doing HF calculation in the lattice limit \cite{das_2020_thesis}.
This explains the experimental results \cite{Li_etal_2019_stm}, where
bond order was observed using Scanning Tunnelling Microscope. This question may also be
important in double layer graphene case, as we might have a similar coexistence. To
understand that possibility one needs to study that lattice Hamiltonian similar
to the one mentioned in Eq. \ref{eq:latHam}. Furthermore, this theory can be used to
explore the
phase diagrams at other filling fractions in the parameter space of $K_{xy},K_z$
and $L_z$ similar to the MLG case \cite{Sodemann_MacDonald_2014}.
There is also a surge of interest in understanding the BCS/BEC condensation
\cite{liu_etal_2020_crossover,Eisenstein_2014,Liu_etal_2017,Li_etal_2017} in
double layer graphene systems. As previously mentioned this state breaks the layer $U(1)$
symmetry just like a superfluid state. At low enough temperatures these excitons
can form a superfluid state where the interaction between the electron and hole can
be tuned by tuning the $L_z$ parameter (which depends on the interlayer separation $d$).

\begin{acknowledgements}
We thank G. Murthy, E. Berg, and J. S. Hofmann for useful discussions. AS would like
to thank the US-Israel Binational Science Foundation for its support via grant no.
2016130 and the University of Kentucky Center for Computational Sciences and
Information Technology Services Research Computing for their support and use of the
Lipscomb Compute Cluster and associated research computing resources. AD was
supported by the German-Israeli Foundation (GIF) Grant No. I-1505-303.10/2019 and
the Minerva Foundation. AD also thanks Weizmann Institute of Science, Israel Deans
fellowship, and Israel planning and budgeting committee for financial support.

\onecolumngrid

\appendix

\section{Details of the technique}
The interacting Hamiltonian in Eq. \ref{eq:contham} can be written in a simplified form as, 

\begin{align}
H_{int}=\frac{\pi l^2}{A}\sum_{\substack{\mathbf{q},k_1,k_2\\a,b,c,d}} e^{-iq_x(k_1-k_2-q_y)l^2-\frac{q^2l^2}{2}}V_{a,b,c,d}:c^\dagger_{k_1-q_y,a} c_{k_1,b} c^\dagger_{k_2+q_y,c} c_{k_2,b}:
\end{align}

where

\begin{align}
V_{a,b,c,d}= K_{xy} &\sum_{i=1,2}  \left(\sigma^0 \otimes \tau^i \otimes P_L\right)_{a,b}\left(\sigma^0 \otimes \tau^i \otimes P_L\right)_{c,d} \nonumber 
+ K_z \left(\sigma^0 \otimes \tau^3 \otimes P_L\right)_{a,b}\left(\sigma^0 \otimes \tau^3 \otimes P_L\right)_{c,d} \\
&+ L_z \left(\sigma^0 \otimes \tau^0 \otimes \gamma^3\right)_{a,b}\left(\sigma^0 \otimes \tau^0 \otimes \gamma^3\right)_{c,d}.
\end{align}

To calculate the total HF energy $E=\langle HF|H_0+H_{int}|HF\rangle$ we write the
average of the four-fermion term that arises in the interacting Hamiltonian $H_{int}$ as

\begin{equation}
\langle HF|c^\dagger_{k_1-q_y,a}c^{\phantom{\dagger}}_{k_1,b}c^\dagger_{k_2+q_y,c}c^{\phantom{\dagger}}_{k_2,d}|HF\rangle = \Delta_{b,a}\Delta_{d,c}\delta_{q_y,0}-\Delta_{d,a}\Delta_{b,c}\delta_{q_y,k_1-k_2}.    
\end{equation}

The first term gives the Hartree term and the second term is the Fock term. Using this, we can calculate the energy from the electron-electron interaction given by $E_{int}=\langle HF|H_{int}|HF\rangle$

\onecolumngrid
\begingroup
\allowdisplaybreaks
\begin{subequations}
\begin{align}    
\nonumber
E_{int}=&\frac{\pi l^2}{A}\sum_{\substack{a,b\\c,d}}\sum_{\substack{\mathbf{q}\\k_1,k_2}} e^{-iq_x(k_1-k_2-q_y)l^2-\frac{q^2l^2}{2}}V_{a,b,c,d}
\left(\Delta_{b,a}\Delta_{d,c}\delta_{q_y,0}-\Delta_{d,a}\Delta_{b,c}\delta_{k_1-q_y,k_2}\right) \\  \nonumber
=&\frac{\pi l^2}{A}\sum_{\substack{a,b,c,d}}V_{a,b,c,d}
\left(\sum_{\substack{q_x\\k_1,k_2}}e^{-iq_x(k_1-k_2)l^2-\frac{q_x^2l^2}{2}}\Delta_{b,a}\Delta_{d,c}
-\sum_{\substack{\mathbf{q},k_1}}e^{-\frac{q^2l^2}{2}}\Delta_{d,a}\Delta_{b,c}\right) \\ \nonumber
=&\frac{1}{2N_{\Phi}}\sum_{\substack{i\\a,b,c,d}} V_{a,b,c,d}
\left(N_{\Phi}^2\Delta_{b,a}\Delta_{d,c}
-\frac{N_{\Phi}A}{(2 \pi)^2}\int_\mathbf{q}d\mathbf{q}e^{-\frac{q^2l^2}{2}}\Delta_{d,a}\Delta_{b,c}\right) \\ 
=&\frac{1}{2N_{\Phi}}\sum_{\substack{a,b,c,d}}V_{a,b,c,d}
\left(N_{\Phi}^2\Delta_{b,a}\Delta_{d,c}
-\frac{N_{\Phi}A}{2 \pi l^2}\Delta_{d,a}\Delta_{b,c}\right) \\ 
E_{int}=&\frac{N_{\Phi}}{2}\sum_{\substack{i\\a,b,c,d}}V_{a,b,c,d}
\left(\Delta_{b,a}\Delta_{d,c}
-\Delta_{d,a}\Delta_{b,c}\right) 
\end{align}
\label{eq:double_HF_int}
\end{subequations}
\endgroup

where $A$ is the area of the system and $N_\Phi=A/(2\pi l^2)$ is the number of guiding
centers in the system. Hence the total energy of the system per guiding center is

\begin{align}
\frac{E}{N_\Phi}=E_z(\sigma_3\otimes \tau_0 \otimes\gamma_0)_{ab}\Delta_{b,a}+\frac{1}{2}\sum_{a,b,c,d}
V_{a,b,c,d}\left(\Delta_{b,a}\Delta_{d,c}-\Delta_{d,a}\Delta_{b,c}\right).
\end{align}

The first term is the Zeeman contribution and the second term comes from the elecron-electron interaction.

\twocolumngrid

\end{acknowledgements}

\bibliography{GraDoubRef}
\end{document}